\documentclass[AMA,STIX1COL,12pt]{WileyNJD-v2}
\usepackage{amsfonts}
\usepackage{amssymb}
\usepackage{mathtools}
\usepackage{microtype}
\usepackage{booktabs}
\usepackage{bbm}
\usepackage{todonotes}
\usepackage{arydshln}
\usepackage{verbatim} 
\usepackage{multirow}
\usepackage{setspace}
\usepackage{subfig}
\usepackage{lineno}
\renewcommand{\thefigure}{\arabic{figure}}
\newcommand{\LB}{\left[\begin{MAT}(r){l}}
\newcommand{\RB}{\\ \end{MAT}\right]}
\articletype{Short communication}%
\begin{document}
\title{Linear-frictional contact model for 3D
       discrete element (DEM) simulations of granular systems}
\author[1]{Matthew R. Kuhn*}
\author[2]{Kiichi Suzuki}
\author[3]{Ali Daouadji}
\authormark{Kuhn, Suzuki \& Daouadji}
\address[1]{Br. Godfrey Vassallo Prof. of Engrg., \orgname{Univ. of Portland}, 
            \country{U.S.A.}}
\address[2]{Former Prof., \orgname{Saitama University}, \orgaddress{\country{Japan}}, ksuzuki@mail.saitama-u.ac.jp}
\address[3]{\orgname{University of Lyon, INSA Lyon, \orgdiv{Laboratoire GEOMAS}}, \orgaddress{EA 7495, Villeurbanne \state{F-69621}, \country{France}}}
\corres{*Donald P. Shiley School of Engineering,
          University of Portland,
          5000 N. Willamette Blvd.,
          Portland, OR, 97203, USA.
          Email: \email{kuhn@up.edu}}
%
%
%
%
%
%
\abstract[Summary]{
The linear-frictional contact model is the most commonly used
contact mechanism for discrete element (DEM) simulations of
granular materials.
Linear springs with a frictional slider are used for modeling
interactions in directions normal and tangential
to the contact surface.
Although the model is simple in two dimensions,
its implementation in 3D faces certain subtle challenges, and
the particle interactions that occur within a single time-step require
careful modeling with a robust algorithm.
The paper details a 3D algorithm that accounts for the changing
direction of the tangential force within a time-step,
the transition from elastic to slip behavior within a time-step,
possible contact sliding during only part of a time-step,
and twirling and rotation of the tangential force during
a time-step.
Without three of these adjustments, errors are introduced in
the incremental stiffness of an assembly.
Without the fourth adjustment,
the resulting stress tensor is not only incorrect, it is no longer a tensor.
The algorithm also computes the work increments during a time-step,
both elastic and dissipative. 
}
\maketitle
%
\section{\large Introduction}
The discrete element method (DEM) is a
numerical technique for simulating the behavior of
granular systems and investigating the grain-scale
mechanics of these systems \cite{Cundall:1979a}.
The method uses
an explicit time integration to update
the position and rotation for each particle
and at each time in a series of time-steps,
requiring the calculation of the grain-to-grain
contact forces of every contact and at every
time-step.
A well-defined, precise, and robust relationship between contact movement and
contact force is essential for DEM codes,
and the most common movement--force relationship, by far,
is the linear--frictional contact.
With this model, the force components that are
normal and tangential to a contact surface are separately
computed.
The normal (compressive)
contact force between two particles, $f^{\text{n},t+\Delta t}$,
at the time $t+\Delta t$
is simply the accumulated overlap $\zeta^{t+\Delta t}$
of the particles' idealized profiles
multiplied by a normal contact stiffness $k^{\text{n}}$.
The change in tangential force, $\Delta \mathbf{f}^{\text{t}}$,
that occurs during the time-step $\Delta t$
is equal to the two particles' relative
tangential movement, vector $\Delta\boldsymbol{\xi}$, during the
time-step multiplied by the tangential
stiffness $k^{\text{t}}$, but the magnitude of the accumulated tangential
force, $|\mathbf{f}^{\text{t},t+\Delta t}|$, is limited to a
friction coefficient $\mu$ multiplied by the normal
force.
These two rules are conventionally written as
\par
\begin{align}
  \label{eq:fn}
  &f^{\text{n},t+\Delta t} = k^{\text{n}}\zeta^{t+\Delta t} \\
  \label{eq:ft}
  &\mathbf{f}^{\,\text{t},t+\Delta t} = \mathbf{f}^{\,\text{t},t}
     + \Delta\mathbf{f}^{\,\text{t}},
     \quad
  \Delta\mathbf{f}^{\,\text{t}} = k^{\text{t}}\Delta\boldsymbol{\xi},
  \quad
  |\mathbf{f}^{\,\text{t},t+\Delta t}| \le \mu f^{\text{n},t+\Delta t}\\
  \label{eq:ftotal}
  &\mathbf{f}^{t+\Delta t} =
  -f^{\text{n},t+\Delta t}\mathbf{n}
  +\mathbf{f}^{\text{t},t+\Delta t}
\end{align}
where $\mathbf{f}^{\,\text{t},t}$
and $\mathbf{f}^{\,\text{t},t+\Delta t}$ are the
tangential forces at the start and end of the time-step, and
$\mathbf{n}$ is the unit vector normal to the contact plane
at the end of the time-step.
The two rules may seem similar,
but they are fundamentally different, with the tangential
rule being the more complex.
The normal force is simply a function of the accumulated overlap
of the particles; whereas,
the tangential rule is
an \emph{incremental} relationship.
The normal force is readily expressed as a scalar;
whereas, the accumulated tangential force $\mathbf{f}^{\text{t}}$
and the tangential increment $\Delta\mathbf{f}^{\text{t}}$
are vectors that lie in the contact's tangential plane but
may have different directions.
These difficulties are particularly complex for three-dimensional
(3D) particles, and
the complexity of the tangential relationship is the subject
of this brief note, which is focused on the ``$\Delta$'' changes
that occur \emph{within a single DEM time-step}.
The drawbacks of conventional frictional models have been
noticed by other authors \cite{Matuttis:2014a}, and 
although subtle, the evolution of $\mathbf{f}^{\text{t}}$
within a time-step affects the full increment
$\Delta\mathbf{f}^{\text{t}}$, an effect that is rarely
included in DEM codes.
\par
In this note, we consider four refinements of the tangential
force in Eq.~(\ref{eq:ft}):
\begin{enumerate}
\item
Fresh contact between two particles can occur within the span of
a single time-step $\Delta t$,
and we correct for any small displacement that occurs before
contact is established.
\item
For an established contact that is not initially slipping but
is slipping at the end of $\Delta t$,
slip does not occur throughout the full time-step,
and we correct for the elastic displacement that precedes slip.
\item
As a contact undergoes slip,
the direction of the tangential force $\mathbf{f}^{\,\text{t}}$
can differ from
the direction of tangential movement $\Delta\boldsymbol{\xi}$,
causing the direction of the tangential force to
change within the time-step.
We correct for this directional change,
which will usually produce a force increment
$\Delta\mathbf{f}^{\text{t}}$ that is not aligned
with the movement $\Delta\boldsymbol{\xi}$.
\item
The two particles can roll and twirl during a time-step,
and we correct for any rigid rotation of the particles
and of the contact force between them.
\end{enumerate}
In this short communication, we develop an algorithm,
detailed in Fig.~\ref{fig:algorithm}, to resolve these
refinements.
We also provide means of calculating the mechanical
work done within a contact, both elastic and dissipated.
Two examples are presented that illustrate the
benefits of applying the
four refinements. 
%
\section{Algorithm}
\begin{figure}
  \centering
  \includegraphics{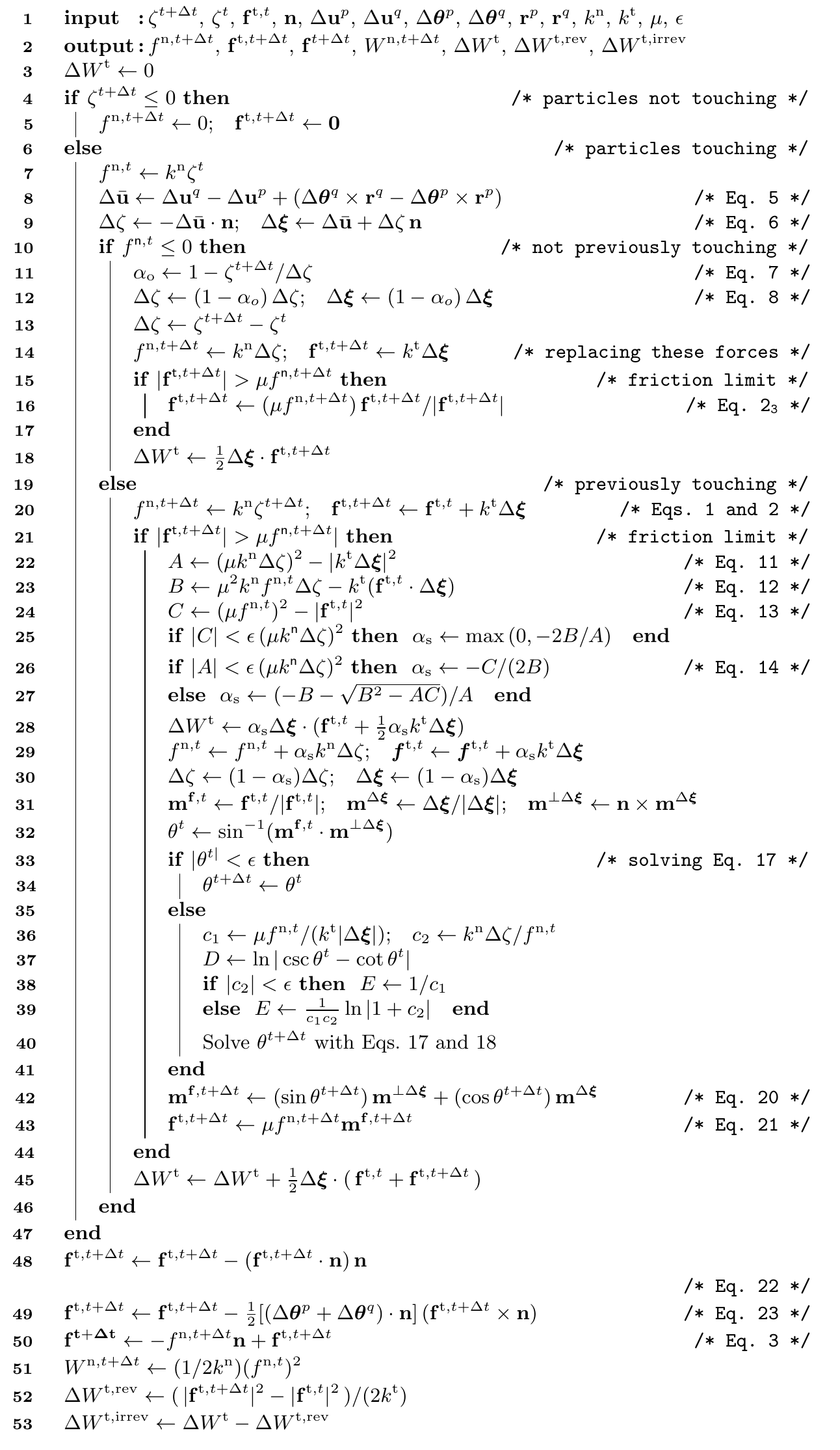} 
  \caption{\ Algorithm for computing the contact force and mechanical work.
           \label{fig:algorithm}}
\end{figure}

Because the refinements listed above apply within the course of a time-step,
we assume that the contact's movements,
$d\zeta$ and $d\boldsymbol{\xi}$, measured from the start of $\Delta t$
are proportional and uniform within $\Delta t$, so that they are
parameterized as 
\begin{equation}\label{eq:alpha}
  d\zeta = \alpha\Delta\zeta \quad\text{and}\quad
  d\boldsymbol{\xi} = \alpha\Delta\boldsymbol{\xi}
\end{equation}
where $\alpha$ proceeds from 0 to 1 during the full
step $\Delta t$.
Note that the normal movement $\zeta$ is a scalar;
whereas, tangential movement $\boldsymbol{\xi}$ is a vector,
although its bold font might be indistinct.
With the assumption of Eq.~(\ref{eq:alpha}),
the paper describes an algorithm that incorporates the four 
refinements listed above.
The algorithm, summarized in Fig.~\ref{fig:algorithm}, requires the following
input information (line~1):
the particles' overlaps at times
$t$ and $t+\Delta t$, the two particles' movements
during $\Delta t$, the contact's stiffness and its friction coefficient,
and other secondary information,
including a small $\epsilon$ that gives the machine precision. 
With DEM, one computes the overlap $\zeta^{t+\Delta t}$
of the particles from the geometric descriptions of their shapes,
and this overlap at $t+\Delta t$ is used to find the
normal force $f^{\text{n},t+\Delta t}$, using Eq.~(\ref{eq:fn})
(lines~7 and~20).
If the contact overlap is negative, the contact force is,
of course, zero (lines~4 and~5).
If the particles are touching, then we
find the normal and tangential
$\Delta$ movements in Eq.~(\ref{eq:alpha})
from the full
contact movement $\Delta \bar{\mathbf{u}}$ of particle $q$
relative to particle $p$ at their contact (lines~8 and~9),
%
\par
\begin{gather}\label{eq:deltau}
  \Delta\bar{\mathbf{u}} = \Delta\mathbf{u}^{q} - \Delta\mathbf{u}^{p}
     + \left(
         \Delta\boldsymbol{\theta}^{q}\times\mathbf{r}^{q}
       - \Delta\boldsymbol{\theta}^{p}\times\mathbf{r}^{p}
       \right)
  \\
  \label{eq:ft1}
  \Delta\zeta =
    -\Delta\bar{\mathbf{u}}\cdot\mathbf{n}
  \quad\text{and}\quad
  \Delta\boldsymbol{\xi} =
    \Delta\bar{\mathbf{u}}
    + \Delta\zeta\,\mathbf{n}
\end{gather}
where the $\mathbf{u}^{p}$, $\mathbf{u}^{q}$,
$\Delta\boldsymbol{\theta}^{p}$, and $\Delta\boldsymbol{\theta}^{q}$
are the particles'
translations and rotations,
$\mathbf{r}^{p}$ and $\mathbf{r}^{q}$ are the contact vectors
that connect the centers of $p$ and $q$ with the contact point,
and $\mathbf{n}$ is the contact's unit normal
vector at the end of the time step, directed outward from $p$. 
Note that a positive incremental overlap $\Delta\zeta$ produces
a compressional normal force.
\par
If the particles are touching at the end of $\Delta t$
but were not touching at the start of the time-step,
then we compute the value $\alpha_{\text{o}}$ at which
the particles first touch (line~11),
\begin{equation}
  \alpha_{\text{o}} = 1-\zeta^{t+\Delta t} / \Delta\zeta
\end{equation}
and we replace $\Delta\zeta$ and $\Delta\boldsymbol{\xi}$
with the movements that occur \emph{after} the particles have touched,
\begin{equation}
  \label{eq:dxi_adjust}
  \Delta\boldsymbol{\xi} \leftarrow (1-\alpha_{\text{o}})\Delta\boldsymbol{\xi}
  ,\quad
  \Delta\zeta \leftarrow (1-\alpha_{\text{o}})\Delta\zeta
\end{equation}
where the arrow ``$\leftarrow$'' represents substitution (assignment)
when coding the algorithm for contact force (line~12).
If the contact is touching at both the start and end of $\Delta t$,
then $\alpha_{\text{o}} = 0$ in Eq.~(\ref{eq:dxi_adjust}).
The values of $\Delta\boldsymbol{\xi}$
and $\Delta\zeta$
are then used with Eq.~(\ref{eq:ft}) to compute a
tangential force increment
$\Delta \mathbf{f}^{\text{t}}$
and the normal force $f^{\text{n},t+\Delta t}$
(lines~13 and~14).
If the particles were not touching at the start of
$\Delta t$, then the full tangential force
$\mathbf{f}^{\text{t}}$ is equal to
the increment $\Delta \mathbf{f}^{\text{t}}$ (line~14).
The tangential force must also be checked to satisfy
the friction limit of Eq.~(\ref{eq:ft}\textsubscript{3}),
as in line~16.
\par
\begin{figure}
  \centering
  \includegraphics{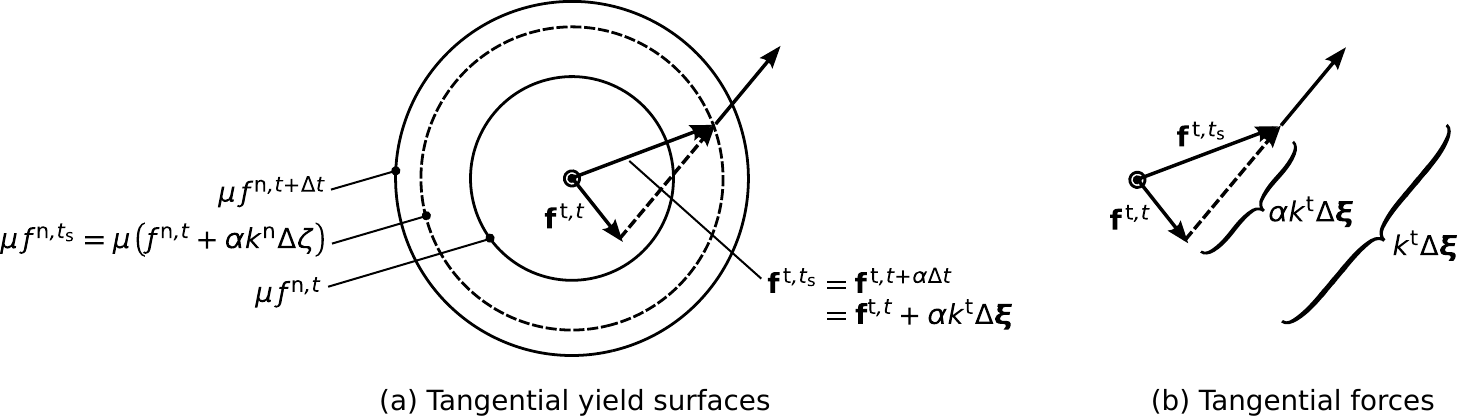}
  \caption{\ Contact tangent plane when
           the initial movement is elastic,
           followed by slip.
           \label{fig:Yield}}
\end{figure}

If the particles were already touching at the start of $\Delta t$,
then the increment $\Delta \mathbf{f}^{\text{t}}$
in Eq.~(\ref{eq:ft}\textsubscript{2})
is merely an estimate of the actual increment,
as described now.
Having established the instant of contact,
we then compute the tangential contact force.
If the magnitude $|\mathbf{f}^{\,\text{t},t+\Delta t}|$
at $t+\Delta t$ exceeds the
friction limit $\mu f^{\text{n},t+\Delta t}$,
then the contact will slide and
we must find the instant $\alpha_{\text{s}}$ at which sliding
commences.
The situation is illustrated in Fig.~\ref{fig:Yield},
which represents a contact's tangent plane, such that the
vectors of
tangential force $\mathbf{f}^{\text{t}}$ 
and movement $\boldsymbol{\xi}$ lie within this plane.
At the start of $\Delta t$ (i.e., when $\alpha=0$),
the limiting magnitude of the tangential force is
$\mu f^{\text{n},t}$, but this limit will change during
the course of $\Delta t$, as $\alpha$ proceeds from 0 to 1,
due to a concurrent change in the normal force
$f^{\text{n}}$ (see Eq.~\ref{eq:alpha}\textsubscript{1}).
The changing
friction limit is represented by circular yield
surfaces in the figure.
The tangential force also changes during $\Delta t$,
and at some point $\alpha_{\text{s}}$,
the tangential force will reach
the friction limit and sliding will commence
(represented by the dashed circle in Fig.~\ref{fig:Yield}).
Within the interval $\alpha\in [0,\alpha_{\text{s}}]$,
the behavior is elastic,
and the tangential force at time $t+\alpha\Delta t$ is given by
$\mathbf{f}^{\text{t},t+\alpha\Delta t}
=\mathbf{f}^{\text{t},t}+\alpha k^{\text{t}}\Delta\boldsymbol{\xi}$,
as in Eq.~(\ref{eq:ft}).
The corresponding friction limit at $t+\alpha\Delta t$ is
$\mu f^{\text{n},t+\alpha\Delta t}=\mu f^{\text{n},t} + \alpha\mu k^{\text{n}}\Delta\zeta$,
as in Eqs.~(\ref{eq:fn}) and~(\ref{eq:alpha}\textsubscript{1}).
To solve for $\alpha_{\text{s}}$ at the instant the tangential
force reaches the friction limit,
we equate the scalar magnitude
$|\mathbf{f}^{\text{t},t+\alpha\Delta t}|$
and the scalar limit $\mu f^{\text{n},t+\alpha\Delta t}$
by applying the law of cosines,
\begin{equation}
  \left(\mu f^{\text{n},t}
  + \alpha_{\text{s}}\mu k^{\text{n}}\Delta\zeta\right)^{2}
  =
  \left|\mathbf{f}^{\text{t},t}\right|^{2}
  +
  \left|\alpha_{\text{s}} k^{\text{t}}\Delta\boldsymbol{\xi}\right|^{2}
  +
  2 \alpha_{\text{s}} k^{\text{t}}
  \left(\mathbf{f}^{\text{t},t}
  \cdot
  \Delta\boldsymbol{\xi}\right)
\end{equation}
where the squared quantity on the left is the radius
of the dashed circle in Fig.~\ref{fig:Yield}.
We solve for $\alpha_{\text{s}}$ (lines~22--27), as
\begin{align}
  \alpha_{\text{s}} &= \left( -B - \sqrt{B^{2}-AC}\right)/A \\
  A &= \left(\mu k^{\text{n}}\,\Delta\zeta\right)^{2}
       - \left|k^{\text{t}}\,\Delta\boldsymbol{\xi}\right|^{2} \\
  B &= \mu^{2} k^{\text{n}} f^{\text{n},t}\,\Delta\zeta
       - k^{\text{t}}\left(\mathbf{f}^{\text{t},t}\cdot\Delta\boldsymbol{\xi}\right) \\
  C &= \left(\mu f^{\text{n},t}\right)^{2} - \left|\mathbf{f}^{\text{t},t}\right|^{2}
\end{align}
Note that if the contact has already reached the friction limit at the
start of $\Delta t$, then $C=0$ and
$\alpha_{\text{s}}$ is the larger of $0$ and $-2B/A$ (line~25).
When $A=0$, we apply L'H\^{o}spital's to find $\alpha_{\text{s}}$
(line~29):
\begin{equation}
  A=0 \;\Rightarrow\; \alpha_{\text{s}}=-\frac{C}{2B}
\end{equation}
In the algorithm of Fig.~\ref{fig:algorithm},
we use a small input parameter $\epsilon$ to test the proximity
of $A$ and $C$ to zero (lines~25 and~26).
\par
Sliding commences at time $t_{\text{s}}=t + \alpha_{\text{s}}\Delta t$,
when the normal and tangential forces,
$f^{\text{n},t_{\text{s}}}$ and $\mathbf{f}^{\text{t},t_{\text{s}}}$,
are
\begin{equation}\label{eq:fts}
  f^{\text{n},t_{\text{s}}}=
  f^{\text{n},t} + \alpha_{\text{s}} k^{\text{n}}\Delta\zeta
  \quad\text{and}\quad
  \mathbf{f}^{\text{t},t_{\text{s}}}
  =
  \mathbf{f}^{\text{t},t} + \alpha_{\text{s}}k^{\text{t}}\Delta\boldsymbol{\xi}
\end{equation}
and for the next set of calculations, we reset $\alpha$ to zero,
and replace $\mathbf{f}^{\text{t},t}$ with $\mathbf{f}^{\text{t},t_{\text{s}}}$,
$f^{\text{n},t}$ with $f^{\text{n},t_{\text{s}}}$,
$\Delta\boldsymbol{\xi}$ with $\alpha_{\text{s}}\Delta\boldsymbol{\xi}$, and
$\Delta\zeta$ with $\alpha_{\text{s}}\Delta\zeta$
(line~30).
\par
Having established the start of sliding,
we proceed to find the tangential force at the end of $\Delta t$.
At the start of sliding,
the tangential force has the unit direction
$\mathbf{m}^{\mathbf{f},t}=
\mathbf{f}^{\text{t},t}/|\mathbf{f}^{\text{t},t}|$.
This direction will likely differ from the
direction
$\mathbf{m}^{\Delta\boldsymbol{\xi}}=\Delta\boldsymbol{\xi}/|\Delta\boldsymbol{\xi}|$
of the tangential movement
during $\Delta t$.
If so, the final direction of the tangential force
$\mathbf{m}^{\mathbf{f},t+\Delta t}$ at time $t+\Delta t$
will \emph{not} coincide with either the initial direction
$\mathbf{m}^{\mathbf{f},t}$, the direction of
the movement $\mathbf{m}^{\Delta\boldsymbol{\xi}}$, or
the direction of the estimated tangential force
$\mathbf{f}^{\,\text{t},t+\Delta t} = \mathbf{f}^{\,\text{t},t}
     + k^{\text{t}}\Delta\boldsymbol{\xi}$,
as given in Eq.~(\ref{eq:ft}).
Rather, sliding will begin in the initial direction
of the tangential force, $\mathbf{m}^{\mathbf{f},t}$,
but this sliding will occur concurrently with the
elastic movement that is orthogonal to the slip direction,
and this elastic movement will alter the force direction
$\mathbf{m}^{\mathbf{f}}$ during the increment $\Delta t$.
The final direction of the tangential force,
$\mathbf{m}^{\mathbf{f},t+\Delta t}$, will be intermediate
between the directions $\mathbf{m}^{\mathbf{f},t}$ and
$\mathbf{m}^{\Delta\boldsymbol{\xi}}$.
\par
\begin{figure}
  \centering
  \includegraphics{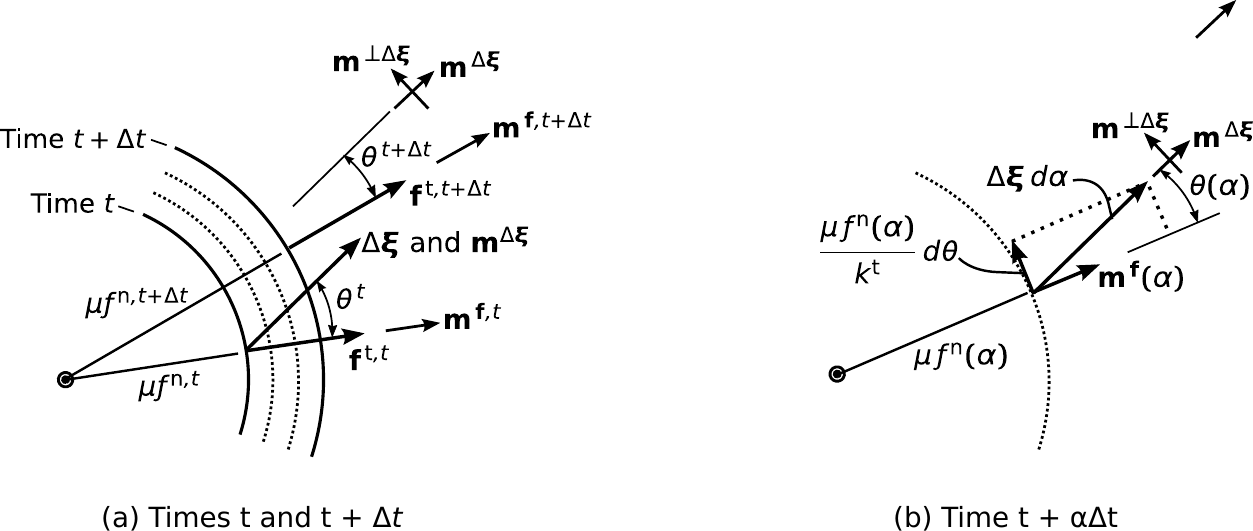}
  \caption{\ Evolution of the tangential force and sliding direction:
           (a)~yield surfaces and the initial and final directions
           of the tangential force, and (b)~increment in movement
           tangent to the yield surface.
           Note that the $\theta$ angles are negative in the
           directions of this figure.
           \label{fig:Theta}}
\end{figure}

This situation is illustrated in Fig.~\ref{fig:Theta},
which represents the tangential plane of the contact.
Recall that we have reset $\alpha$ to 0 and have
reset $\mathbf{f}^{\text{t},t}$, $f^{\text{n},t}$,
and $\Delta\boldsymbol{\xi}$ to their values at the start 
of sliding, $t_{\text{s}}$.
The tangential plane of the figure
contains the tangential force $\mathbf{f}^{\text{t},t}$ at the
start of sliding (the $\mathbf{f}^{\text{t},t_{\text{s}}}$ in
Eq.~\ref{eq:fts});
the tangential movement $\Delta\boldsymbol{\xi}$;
and the final tangential force
$\mathbf{f}^{\,\text{t},t+\Delta t}$.
The unit vectors $\mathbf{m}^{\mathbf{f},t}$
and $\mathbf{m}^{\Delta\boldsymbol{\xi}}$ lie within
this plane, with
$\mathbf{m}^{\mathbf{f},t}=\mathbf{f}^{\text{t},t}/|\mathbf{f}^{\text{t},t}|$
aligned with
the \emph{initial} force $\mathbf{f}^{\text{t},t}$, and
$\mathbf{m}^{\Delta\boldsymbol{\xi}}
=\Delta\boldsymbol{\xi}/|\Delta\boldsymbol{\xi}|$ is
aligned with $\Delta\boldsymbol{\xi}$.
We also define a unit vector
$\mathbf{m}^{\perp\Delta\boldsymbol{\xi}}$ that is
perpendicular to the direction of
tangential movement $\mathbf{m}^{\Delta\boldsymbol{\xi}}$
(the vector $\mathbf{m}^{\perp\Delta\boldsymbol{\xi}}$ can be
computed with the cross-product
$\mathbf{m}^{\perp\Delta\boldsymbol{\xi}}=\mathbf{n}\times \mathbf{m}^{\Delta\boldsymbol{\xi}}$,
where $\mathbf{n}$ is the contact's normal direction
at the end of the time-step, line~31).
At the start of sliding, when $\alpha=0$,
direction $\mathbf{m}^{\Delta\boldsymbol{\xi}}$ makes
angle $\theta^{t}$ with the
initial force direction $\mathbf{m}^{\mathbf{f},t}$,
such that
$\theta^{t}=\sin^{-1}(\mathbf{m}^{\perp\Delta\boldsymbol{\xi}}\cdot\mathbf{m}^{\mathbf{f},t})$,
as in line~32.
Although the directions $\mathbf{m}^{\Delta\boldsymbol{\xi}}$
and $\mathbf{m}^{\perp\Delta\boldsymbol{\xi}}$
do not change during the time-step
(recall Eq.~\ref{eq:alpha}), 
the force orientation $\mathbf{m}^{\mathbf{f}}$
and its corresponding angle $\theta$ will change,
with $\theta$ starting at $\theta^{t}$
and ending at $\theta^{t+\Delta t}$,
and we now describe the method for finding this
final angle.
The circle in Fig.~\ref{fig:Theta}a is a yield surface that defines
the frictional limit $\mu f^{\text{n}}$ of the tangential force.
Any tangential contact movement that is tangent to this circle
(that is, movement within the tangent plane that is
perpendicular to the current tangential
force direction, $\mathbf{m}^{\mathbf{f}}$)
is elastic.
A change in the angle $\theta$ by the increment $d\theta$
corresponds to an increment of elastic movement
perpendicular to $\mathbf{m}^{\mathbf{f}}$,
equal to the tangential force increment divided by the
elastic stiffness $k^{\text{t}}$ (Fig.~\ref{fig:Theta}b).
This elastic force increment also equals the radius of the
yield surface, $\mu f^{\text{n}}$, multiplied by $d\theta$.
Noting that direction $\mathbf{m}^{\mathbf{f}}$
makes angle $\theta$ with $\mathbf{m}^{\Delta\boldsymbol{\xi}}$
(Fig.~\ref{fig:Theta}b, where the $\theta$ are shown as negative
in this figure),
the magnitude of the full increment of movement
$|\Delta\boldsymbol{\xi}\,d\alpha|$ during increment $d\alpha$
is given by
\begin{equation}\label{eq:dalpha}
  |\Delta\boldsymbol{\xi}\,d\alpha|
  = -\frac{\mu f^{\text{n}}(\alpha)}{k^{\text{t}}\sin\theta}\,d\theta
\end{equation}
%
%
where $\alpha$ proceeds from 0 to 1.
Note that angle $\theta$ will be \emph{reduced} during the time-step,
as $\mathbf{m}^{\mathbf{f}}$ becomes more aligned
with $\mathbf{m}^{\Delta\boldsymbol{\xi}}$,
which is consistent with the negative in the equation.
In this equation, the normal force $f^{\text{n}}$
will change during time-step $\Delta t$ and, as such,
is a function of $\alpha$
(see Eqs.~\ref{eq:fn} and~\ref{eq:alpha}), 
\begin{equation}
f^{\text{n}}(\alpha) =
f^{\text{n},t} + \alpha k^{\text{n}}\Delta\zeta
\end{equation}
%
Equation~(\ref{eq:dalpha}) is a first-order separable
differential equation on the domain $\alpha\in[0,1]$, in which
angle $\theta$ starts at $\theta^{t}$ and ends at
$\theta^{t+\Delta t}$.
The solution is (lines~36--40)
\begin{gather}\label{eq:thetadt}
  \ln \left|
  \csc\theta^{t+\Delta t} - \cot\theta^{t+\Delta t}
  \right| =
  \ln \left|
  \csc\theta^{t} - \cot\theta^{t}
  \right|
  -
  \frac{1}{c_{1}c_{2}}
  \ln | 1 + c_{2} | \\
  \label{eq:signs}
  \theta^{t+\Delta t} \leftarrow
  \text{sgn}(\theta^{t})\,\theta^{t+\Delta t}
\end{gather}
where the parameters $c_{1}$ and $c_{2}$ are as follows:
$c_{1}=\mu f^{\text{n},t}/(k^{\text{t}}|\Delta\boldsymbol{\xi}|)$
and $c_{2}=k^{\text{n}}\Delta\zeta / f^{\text{n},t}$.
Note that the final term on the right, $(1/c_{1}c_{2})\ln (1+c_{2})$,
is replaced with $1/c_{1}$ when $c_{2}=0$ (line~38).
When $\theta^{t}=0$, the direction of movement $\Delta\boldsymbol{\xi}$
is aligned with the initial force $\mathbf{f}^{\text{t},t}$, and
$\theta^{t+\Delta t}=\theta^{t}$ (line~34).
The $\theta^{t+\Delta t}$ that is computed with Eq.~(\ref{eq:thetadt})
will always be positive, so we must intervene so that its sign
conforms with that of $\theta^{t}$,
as in Eq.~(\ref{eq:signs}).
\par
Solving $\theta^{t+\Delta t}$ with the nonlinear Eq.~(\ref{eq:thetadt}) can be
computationally taxing,
and rather than using Newton's method (or another iterative approach),
we can simply create a look-up table and interpolate to determine
$\ln |\csc\theta-\cot\theta|$ to find the term on the right,
and we can create an inverse look-up table and interpolate for
$(\csc\theta -\cot\theta)$ to determine $\theta^{t+\Delta t}$.
These master tables are efficiently reused with every
contact and at every time-step.
Once $\theta^{t+\Delta t}$ is computed, we find the direction of
the final tangential force, and its full value is
the vector sum of its $\mathbf{m}^{\mathbf{m}\perp,t}$
and $\mathbf{m}^{\mathbf{m},t}$ components (lines~42--43),
\begin{align}\label{eq:ms}
  \mathbf{m}^{\mathbf{f},t+\Delta t} &=
    (\sin\theta^{t+\Delta t})\mathbf{m}^{\perp\Delta\boldsymbol{\xi}} +
    (\cos\theta^{t+\Delta t})\mathbf{m}^{\Delta\boldsymbol{\xi}}\\
  \label{eq:ftnew}
  \mathbf{f}^{\text{t},t+\Delta t} &=
    \mu f^{\text{n},t+\Delta t}
    \mathbf{m}^{\mathbf{f},t+\Delta t}
\end{align}
where we have applied the friction limit on magnitude
$|\mathbf{f}^{\text{t},t+\Delta t}|$,
as in Eq.~(\ref{eq:ft}\textsubscript{3}).
\par
\begin{figure}
  \centering
  \includegraphics{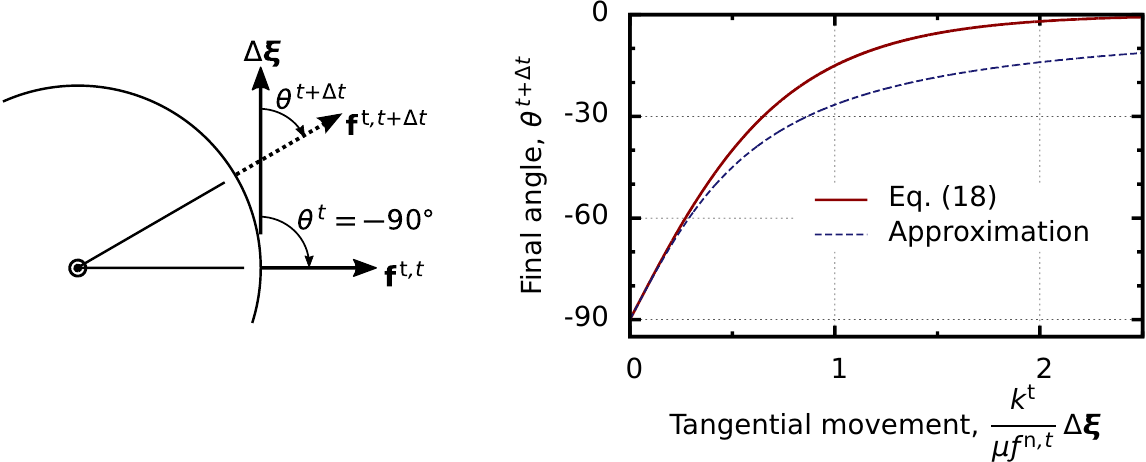}
  \caption{\ Example showing the effect of the magnitude of
           the tangential movement $\Delta\boldsymbol{\xi}$
           on the final direction of the tangential
           force.  The movement $\Delta\boldsymbol{\xi}$
           is applied orthogonal to the initial
           force $\mathbf{f}^{\text{t},t}$.
           \label{fig:example}}
\end{figure}

Figure~\ref{fig:example} shows the effect of the movement magnitude
$|\Delta\boldsymbol{\xi}|$ on the final direction of the
tangential force.
In this example, the movement vector $\Delta\boldsymbol{\xi}$
is applied orthogonal to the initial tangential
force $\mathbf{f}^{\text{t},t}$, with $\theta^{t}=-90^{\circ}$,
and the normal force $f^{\text{n}}$ is assumed to remain
constant during the time-step $\Delta t$.
Under these conditions, the tangential force
$\mathbf{f}^{\text{t}}$ will rotate as the movement increases,
to become more closely aligned with the direction
of $\Delta\boldsymbol{\xi}$.
When the movement vector $\Delta\boldsymbol{\xi}$ is very small,
the direction of $\mathbf{f}^{\text{t}}$ is barely altered,
and the final angle $\theta^{t+\Delta t}$ remains
near $-90^{\circ}$.
Conversely, with large movements, the force $\mathbf{f}^{\text{t}}$
rotates to conform with the movement direction, with
$\theta^{t+\Delta t}$, approaching $0^{\circ}$.
The figure also shows the commonly used approximation
of $\mathbf{f}^{\text{t},t+\Delta t}$,
in which the increment
$k^{\text{t}}\Delta\boldsymbol{\xi}$ is imply added
to the initial force
$\mathbf{f}^{\text{t},t}$ and then projected onto
the final yield surface to find the final force,
ignoring the progressive change in the direction of slip
during $\Delta t$, as in Eq.~(\ref{eq:dalpha}).
Although the approximation works well when the increment
$\Delta\boldsymbol{\xi}$ is small compared with the
current normal force
(i.e., small compared with the quotient $f^{\text{n}}/k^{\text{t}}$),
large errors occur when the force increment is larger than 20\% of
the initial tangential force.
\par
Having found the new tangential force
$\mathbf{f}^{\text{t},t+\Delta t}$ in the manner
of Eqs.~(\ref{eq:ms}) and~(\ref{eq:ftnew}), 
we must remember that the two particles are moving and rotating 
during time-step $\Delta t$, thus rotating the tangent plane.
As such, the tangential force can drift from its
tangent plane after a series of time-steps.
The fourth (and final) refinement of the force calculation
accounts for this directional drift and for twirling
of the tangential force
within the tangent plane.
In DEM codes,
the drift is usually prevented by
projecting the new force $\mathbf{f}^{\text{t},t+\Delta t}$
onto the new tangent plane \cite{Lin:1997a,Vuquoc:2000a}
(line~48).
Less common, however, is an adjustment that
must be applied \emph{within} the tangent plane.
If the two particles rotate (twirl) about their contact normal $\mathbf{n}$
as a rigid pair, then the tangential force will also
rotate, even in the absence of a relative tangential
movement $\Delta\boldsymbol{\xi}$ between the particles.
This rotation of the tangential force is consistent with
the principle of objectivity, as no rotation
would be seen by an observer who rotates in unison with
the twirling pair;
whereas, a stationary observer would see an equal rotation
of the two particles and of the tangential force.
The projected adjustment and the
the rotated (twirled) adjustment are computed successively
as (lines~48 and~49)
\begin{align}\label{eq:project}
  \mathbf{f}^{\text{t},t+\Delta t}
  & \leftarrow
  \mathbf{f}^{\text{t},t+\Delta t}
  - \left(
    \mathbf{f}^{\text{t},t+\Delta t}\cdot\mathbf{n}
  \right)
  \cdot\mathbf{n}
  \\
  \label{eq:twirl}
  \mathbf{f}^{\text{t},t+\Delta t}
  & \leftarrow
  \mathbf{f}^{\text{t},t+\Delta t}
  -
  \frac{1}{2}
  \left[\rule{0ex}{2ex}
    \left(
      \Delta\boldsymbol{\theta}^{p} + \Delta\boldsymbol{\theta}^{q}
    \right)\cdot\mathbf{n}
  \right]
  \left( \mathbf{f}^{\text{t},t+\Delta t}\times\mathbf{n}\right)
\end{align}
where $\Delta\boldsymbol{\theta}^{p}$ and $\Delta\boldsymbol{\theta}^{q}$
are the particles' incremental rotation vectors
(as in Eq.~\ref{eq:deltau}),
and the twirling is simply the average
of the two rotations about the contact normal $\mathbf{n}$.
\par
As two particles interact, elastic energy is stored or released
from their contact, and energy can also be dissipated in friction.
We are often interested in the micro-scale destination of the
mechanical work that is delivered to a contact.
In Fig.~\ref{fig:algorithm}, we compute the total, accumulated work
$W^{\text{n}}$ done by the normal force and the work increment
$\Delta W^{\text{t}}$ done by the tangential force.
The normal work is entirely elastic,
and its total stored energy can be computed directly from the
final force, as $(f^{\text{n},t+\Delta t})^{2}/(2k^{\text{n}})$,
in line~51.
The tangential
work is computed as an increment
during $\Delta t$ and is the sum
of elastic (reversible) and dissipated (irreversible)
parts,
$\Delta W^{\text{t,rev}}$ and $\Delta W^{\text{t,irrev}}$.
The reversible tangential increment $\Delta W^{\text{t,rev}}$ 
in line~52 simply depends on the
initial and final forces,
but the increment of total tangential work depends upon the
tangential
force--movement path, which is quite complex for a sliding
contact
(this complexity is partly expressed in Eq.~\ref{eq:dalpha},
which can be used to find
the relationship between movement $d\boldsymbol{\xi}$
and the evolving tangential force $\mathbf{f}^{\text{t}}$).
For tangential movements that precede slip,
the work increment $\Delta W^{\text{t}}$ is simply
equal to the elastic increment $\Delta W^{\text{t,rev}}$
(lines~18 and~28).
Once sliding commences, we approximate the tangential
work as the inner product of the movement $\Delta\boldsymbol{\xi}$
and the average of the tangential forces at
the start of sliding and at the end of $\Delta t$
(line~45).
The increment of dissipation $\Delta W^{\text{t,irrev}}$
is equal to the total tangential work $\Delta W^{\text{t}}$
minus the elastic increment $\Delta W^{\text{t,rev}}$
(lines~52--53).
\par
Viscous dissipation is typically used as a means of stabilizing
DEM simulations, and Cundall and Strack \cite{Cundall:1979a}
describe two forms of such damping:
mass-damping of a particle body, which is proportional
to the particle's velocity,
and contact-damping, which is proportional to
the contact velocity $d\bar{\mathbf{u}}/dt$ (see Eq.~\ref{eq:deltau}).
If contact-damping is used, it is effected by simply adding
a damping force $\nu\Delta\bar{\mathbf{u}}/\Delta t$
to the contact force $\mathbf{f}^{t+\Delta t}$ in line~50
of Fig.~\ref{fig:algorithm},
where $\nu$ is the damping constant.
\section{Examples}
Two examples are presented:
the first illustrates the effect of
making the first three adjustments to the conventional
linear--frictional algorithm;
whereas, the second example illustrates
the effects of including the
rolling and twirling of tangential forces
(see the four refinements outlined in the Introduction).
We should emphasize, however, that these improvements will
not appreciably affect certain types of DEM results,
but can become significant when DEM simulations are used for
studying other elements of granular behavior.
The macro-scale behavior over large spans of strain is fairly insensitive
to the contact details: for example, similar strength
results are obtained
from both DEM and Contact Dynamics (CD) simulations at large strains.
The two simulation methods are quite different, and the
latter altogether neglects contact stiffness and
treats the contacts as rigid-frictional.
As another illustration, the macro-scale behavior
of 2D disks and 3D spheres are qualitatively similar,
and one might not even guess a simulation's dimensionality
by looking
at the stress-strain response across large spans of strain.
The effect of the first three linear--frictional refinements
will become more pronounced for
small spans of strain, as with stress-probe studies, and under
conditions in which many contacts are either sliding or
are undergoing combined sliding and elastic movement.
These conditions are explored in the next paragraph, and
the effect of the fourth refinement is addressed further below.
\par
A straightforward simulation of triaxial compression
was conducted on a 3D assembly
of 10,648 smooth non-convex sphere-clusters
contained within periodic boundaries.
The dense initial particle arrangement was isotropic with an
initial porosity 0.363 (void ratio of 0.570),
an inter-particle friction coefficient
$\mu=0.55$, and equal normal and tangential contact
stiffnesses, $k^{\text{n}}=k^{\text{t}}$.
The preliminary stage of loading was drained isobaric
(constant-$p$) triaxial compression in the $x_{1}$ direction,
using the refined model
of this paper.
At peak strength, the ratio of deviator stress $q$ and
mean stress $p$ was about 1.7, but our
example was conducted at a smaller strain,
when $q/p$ was about 0.8, where we suspended the loading
and conducted two series of stress-probes:
one series with the conventional unimproved linear-frictional model
and the other series with the model in the paper.
At this smaller strain, the material was in the early stage
of strain hardening and about 11\% of the contacts were 
either sliding or with a mobilized friction within 0.001\%
of $\mu$.
\par
\begin{figure}
  \centering 
  \includegraphics{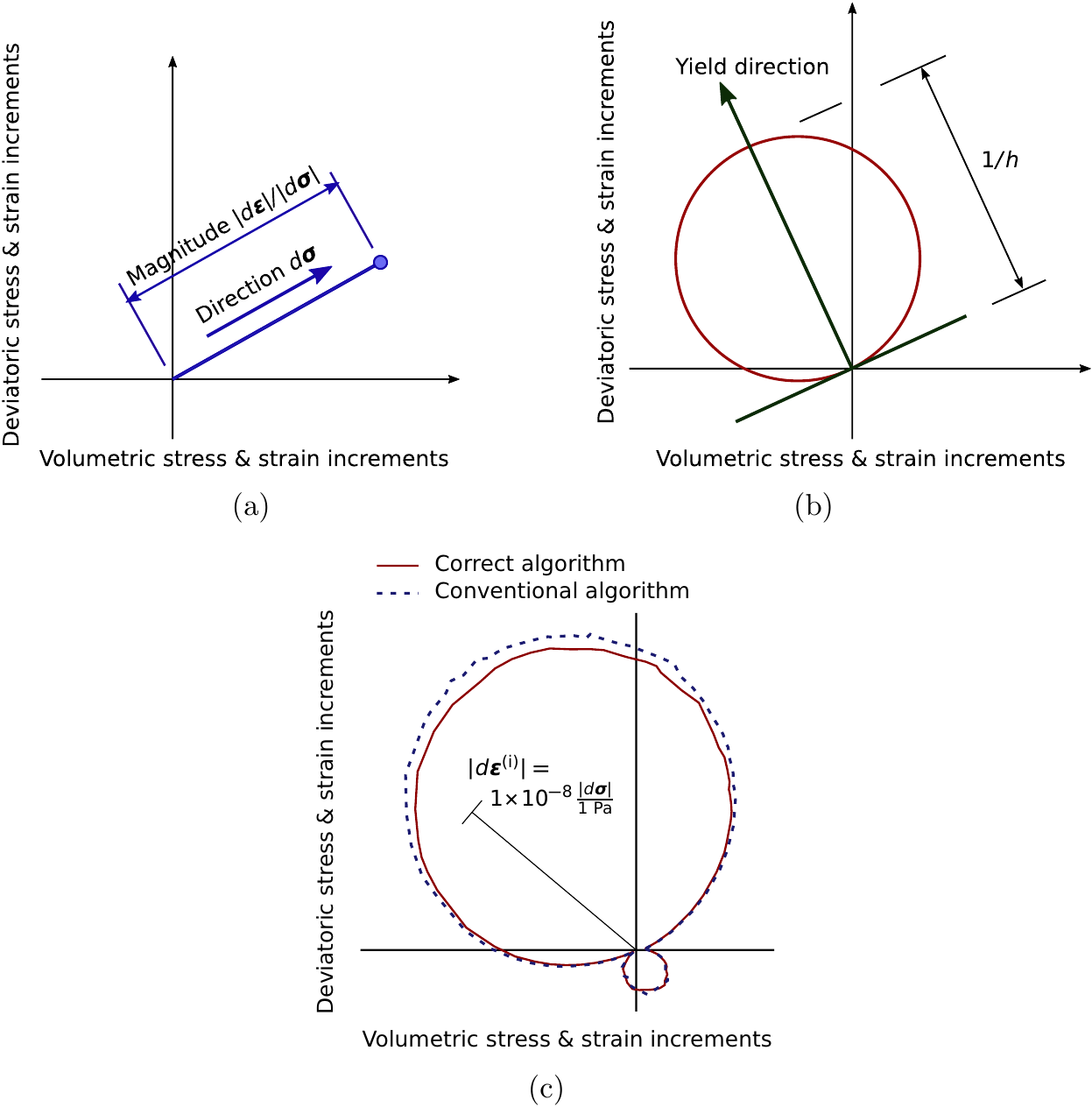}%
\caption{\ Plots of incremental plastic strains from stress-probes
         in the Rendulic plane of volumetric and deviatoric stress
         and strain:
         (a)~technique for plotting the plastic increment of a
         single probe;
         (b)~locus of probe points, as predicted by
         elasto-plastic theory; and
         (c)~locus of probe points for two the linear--frictional
         algorithms.
         \label{fig:circles}}
\end{figure}

For each of the two linear--frictional algorithms,
over 70 axisymmetric
probes were conducted with a technique
previously described by the authors
\cite{Kuhn:2018c},
such that both elastic and plastic strain increments were
measured for a suite of incremental loading directions
(more precisely, reversible and irreversible increments).
All probes were axisymmetric and conducted within the $p$-$q$ Rendulic plane,
with strain-probes of magnitude $6\times 10^{-6}$.
We focused upon the plastic increments for each probe, which we
processed and plotted in a manner that permits direct evaluation
of the yield surface, flow direction, and plastic modulus
as well as  conformance with conventional elasto-plasticity.
The plotting technique is more extensively described elsewhere by the
authors \cite{Kuhn:2018c} and is illustrated in Fig.~\ref{fig:circles}.
The result of a single probe is represented by a single point in
the Rendulic plane of
volumetric and deviatoric strains and stresses,
with the axes normalized to a common orthonormal basis
(volumetric strain $=\varepsilon_{kk}/\sqrt{3}$, and
deviatoric strain $=\sqrt{2/3}(\varepsilon_{11}-\varepsilon_{33})$,
with $\varepsilon_{22}=\varepsilon_{33}$).
Each probe-point is located along a direction that is aligned
with the stress increment, and the radial distance from the origin
is equal to the magnitude of the resulting plastic strain increment divided by
the magnitude of the stress increment (Fig.~\ref{fig:circles}a).
Although this form of plotting is unusual, it presents information
about both the strain and stress increments, and
it provides the following information
in a compact form (Fig.~\ref{fig:circles}b):
(i)~the locus of probe-points will form a single
circle in the Rendulic plane that passes through the origin 
(and a sphere in the full
$\varepsilon_{11}$--$\varepsilon_{22}$--$\varepsilon_{33}$ space),
provided that the material conforms with conventional elasto-plasticity theory;
(ii)~the diameter of the circle is the inverse of the plastic modulus;
and
(iii)~the orientation of the circle equals the direction of the yield surface.
The plot can also include the flow-direction for each stress-direction,
although we did not use this feature with the simplified plots of the paper.
Here we use the plot to compare the hardening moduli from the
series of DEM simulations for the two
linear--frictional contact algorithms.
\par
Figure~\ref{fig:circles}c shows the results for both the conventional
linear--frictional algorithm and that of the paper. 
Each locus includes over 70 points, and each point is the result of
a DEM stress-probe in which the strain increment had a magnitude of
$6\times 10^{-6}$ that was achieved with 30 DEM time-steps.
Each locus is shown to be a circle within the Rendulic plane,
although this is not the case when 
plotted in the deviatoric plane (see the authors' work\cite{Kuhn:2018c},
indicating a failure of elasto-plastic principles when
applied to granular materials).
The tilt of the larger circle indicates the yield direction, and
the presence of a small complementary circle
simply means that, contrary to elasto-plasticity principles, plastic strains
occur in the ``unloading'' direction (within the presumed elastic region),
a result that has been demonstrated by the authors \cite{Kuhn:2018c}
for strains as small as $2\times 10^{-6}$.
\par
The results in the figure show that the hardening modulus
(inverse of the larger circle's diameter) for
the conventional linear--frictional algorithm has an error of
about 5\%, when compared with the paper's algorithm. 
Although small, the error will increase if probes are done with
fewer than 30 time-steps, and
the error will also increase
as loading proceeds toward the peak stress, bringing
more contacts to the friction limit.
\par
\begin{figure}
  \centering
  \includegraphics{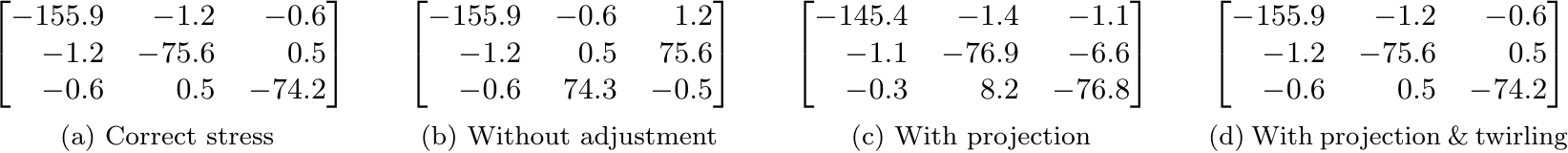}
  \caption{\ Stress tensors computed for assembly of
           10,648 particles that has been loaded with
           constant-$p$ triaxial compression in the $x_1$
           direction.  The units are kPa.
           Subfigure~(a) gives the original tensor $\boldsymbol{\sigma}$.
           The three other tensors have been restored to the
           original frame after a 90$^{\circ}$ rotation about
           the $x_1$ axis:
           (b)~stress $\boldsymbol{\sigma}^{0^{\circ}}$
           when neither the projection nor twirling adjustments
           are made during the 90$^{\circ}$ rotation,
           (c)~stress $\boldsymbol{\sigma}^{0^{\circ}}$
           when only the projection adjustments
           are made during the 90$^{\circ}$ rotation, and
           (d)~stress $\boldsymbol{\sigma}^{0^{\circ}}$
           when both the projection nor twirling adjustments
           are made.
           \label{fig:rotate}}
\end{figure}

We now consider the effect of the fourth
adjustment that is listed in the Introduction
and accomplished with Eqs.~(\ref{eq:project}--(\ref{eq:twirl}).
After computing a new tangential force that results from
the contact movements $\Delta\zeta$ and $\Delta\boldsymbol{\xi}$,
the tangential force must be adjusted in two ways:
(a)~by projecting each contact's force onto its tangent plane and
(b)~by twirling the force about the contact normal
as a result of any co-rotations of the two particles.
Only the projected force is typically computed \cite{Lin:1997a}.
The importance of both adjustments, however,
is illustrated in a simple example.
As with the previous example,
the assembly of 10,648 particles was loaded
in constant-$p$ triaxial compression to a $q/p$ of about
0.8 by compressing the assembly in the $x_{1}$ direction.
The simulation was then stopped at this
stress  $\boldsymbol{\sigma}$, and the entire assembly
was rotated as a rigid body about the $x_{1}$ axis in a sequence
of 1000 rotation increments to an cumulative rotation of
90$^{\circ}$.
Because this rigid rotation produces no relative movements
among the contacts,
neither the $\Delta\zeta$ nor $\Delta\boldsymbol{\xi}$ movements,
the tangential contact forces do not change, even though the forces
must rotate with the entire assembly
(stated differently, an observer who rotates
with the assembly would see no change in the forces,
although a stationary observer would watch the forces rotating).
This rotation is not considered in
Eqs.~(\ref{eq:fn})--(\ref{eq:ftnew}),
but is realized with the adjustments of
Eqs.~(\ref{eq:project}) and~(\ref{eq:twirl}).
The stress after the 90$^{\circ}$
rotation, $\boldsymbol{\sigma}^{90^{\circ}}$,
was computed from the contact data
with the usual Love--Weber
(Piola) summation of dyads of branch vectors and contact forces.
To illustrate the effects of the projection and twirling adjustments,
we restored the stress tensor
to the original (pre-rotated) frame, designated as stress
$\boldsymbol{\sigma}^{0^{\circ}}$,
by applying the tensor transformation
\begin{equation}
  \sigma^{0^{\circ}}_{ij} =
  \Omega^{-90^{\circ}}_{ki}
  \sigma^{90^{\circ}}_{kl}
  \Omega^{-90^{\circ}}_{lj}
\end{equation}
where the rotation tensor $\boldsymbol{\Omega}^{-90^{\circ}}$
effects a full counter-rotation of 90$^{\circ}$.
Standard DEM codes will correctly account for rotations of
the branch vectors as well as rotations of the normal forces,
but a proper algorithm must also rotate the tangential forces.
If correct, the rotated (and then counter-rotated) stress
$\boldsymbol{\sigma}^{0^{\circ}}$
will coincide with the original stress tensor $\boldsymbol{\sigma}$.
Figure~\ref{fig:rotate} shows four stress tensors:
(a)~the original stress  $\boldsymbol{\sigma}$
before the 90$^{\circ}$ rotation, 
(b)~the stress $\boldsymbol{\sigma}^{0^{\circ}}$
when neither the projection nor twirling adjustments
are made during the 90$^{\circ}$ rotation,
(c)~the stress $\boldsymbol{\sigma}^{0^{\circ}}$
when only the projection adjustments
are made during the 90$^{\circ}$ rotation, and
(d)~the stress $\boldsymbol{\sigma}^{0^{\circ}}$
when both the projection and twirling adjustments
are made. 
The results show that
unless both adjustments are made,
the original stress $\boldsymbol{\sigma}$ is not
recovered, and a substantial error results.
Indeed, borrowing a phrase of Wolfgang Pauli,
the stress tensors of Figs.~\ref{fig:rotate}b and \ref{fig:rotate}c,
which do not include the proper adjustments,
``are not even wrong'':  they are not even tensors. 
\section{Discussion}
In the paper, we document an algorithm for computing
the contact force of a linear-frictional contact, which is the simplest
and most common model used in DEM simulations.
The algorithm resolves several subtle difficulties that arise when using
a finite time-step $\Delta t$ but are typically not addressed in
conventional codes.
The first three difficulties involve incremental changes in the tangential
force and lead to differences between the refined approach of the paper
and the more conventional approach of simply projecting an
approximate final force $\mathbf{f}^{\text{t},t+\Delta t}$ onto
the slip (yield) surface.
A stress-probe example
shows that modest errors result in an assembly's incremental stiffness
when the three adjustments are not made.
The error can certainly be reduced by using smaller (and more) time-steps,
but at the expense of using larger probes and additional computation time.
Unless it is properly resolved, the fourth difficulty produces
significant accumulating
errors in the tangential forces, errors that are unaffected by the size
of the time-step.
Without applying the fourth adjustment,
the resulting stress tensor is not only incorrect, it is no longer a tensor.
Although these are practical and compelling reasons for using
an improved algorithm, a more thoughtful argument is that a numerical simulation,
which is intended to model or understand a physical process,
should be faithful to the underlying physics of the process.
%
\bibliographystyle{wileyNJD-APA}
%

\end{document}